\pdfoutput=1
\documentclass[11pt]{article}

\usepackage[final]{acl}

\usepackage{times}
\usepackage{latexsym}
\usepackage{adjustbox}
 \definecolor{mintgreen}{RGB}{24,180,170}

\usepackage[T1]{fontenc}
\usepackage[utf8]{inputenc}

\usepackage{microtype}

\usepackage{inconsolata}

\usepackage{graphicx}

\usepackage{enumitem}
\usepackage{xcolor}
\usepackage{multirow}
\usepackage{booktabs}
\usepackage{bm}
\usepackage{adjustbox}
\usepackage[ruled,linesnumbered,noend]{algorithm2e}
\usepackage{subcaption}
\usepackage{amsmath}
\usepackage{amssymb}

\title{Explain then Rank: Scale Calibration~of~Neural~Rankers Using~Natural~Language~Explanations from LLMs}

\author{
 \textbf{Puxuan Yu\thanks{The work was done while Puxuan was an intern at Dataminr and prior to joining Snowflake.}\textsuperscript{1}},
 \textbf{Daniel Cohen\textsuperscript{2}},
 \textbf{Hemank Lamba\textsuperscript{2}},
 \textbf{Joel Tetreault\textsuperscript{2}},
 \textbf{Alex Jaimes\textsuperscript{2}}
\\
 \textsuperscript{1}Snowflake Inc.
 \textsuperscript{2}Dataminr Inc.
\\
 \small{
   \textbf{Correspondence:} \href{mailto:pxyuwhu@gmail.com}{pxyuwhu@gmail.com}
 }
}

\begin{document}
\maketitle
\begin{abstract}
In search settings, calibrating the scores during the ranking process to quantities such as click-through rates or relevance levels enhances a system's usefulness and trustworthiness for downstream users. While previous research has improved this notion of calibration for low complexity learning-to-rank models, the larger data demands and parameter count specific to modern neural text rankers produce unique obstacles that hamper the efficacy of methods intended for the learning-to-rank setting.

This paper proposes exploiting large language models (LLMs) to provide relevance and uncertainty signals for these neural text rankers to produce scale-calibrated scores  through Monte Carlo sampling of natural language explanations (NLEs). Our approach transforms the neural ranking task from ranking textual query-document pairs to ranking corresponding synthesized NLEs. Comprehensive experiments on two popular document ranking datasets show that the NLE-based calibration approach consistently outperforms past calibration methods and LLM-based methods for ranking, calibration, and query performance prediction tasks.
\end{abstract}

\section{Introduction}

\begin{figure*}
    \centering
    \includegraphics[width=0.7\linewidth]{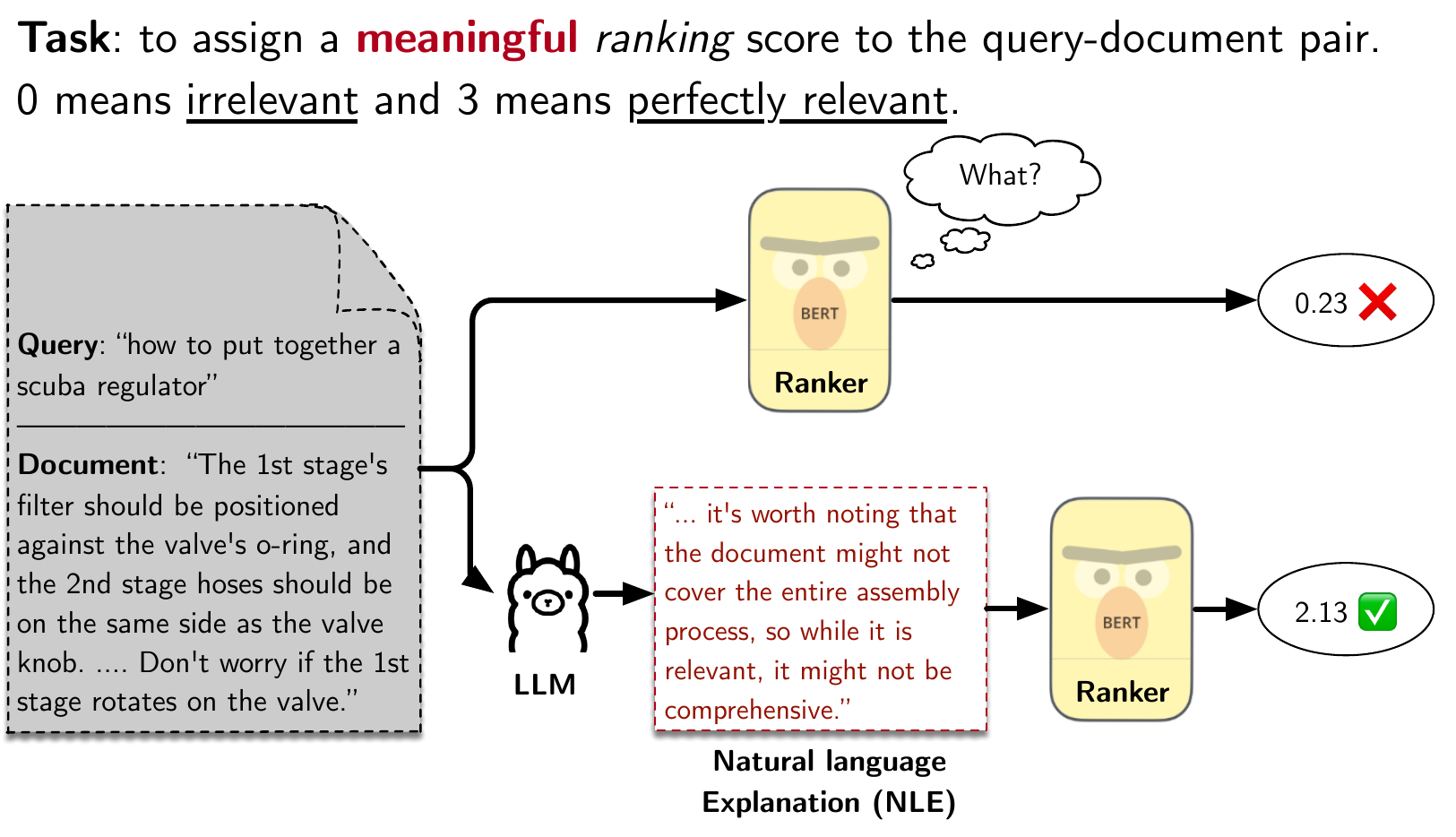}
    \caption{The key idea of this study: Neural ranking models struggle to produce {\em meaningful} ranking scores when encountering complex query-document pairs. We investigate the integration of natural language explanations as inputs to neural rankers, aiming to simplify the scale-calibrated ranking task for these rankers.}
    \label{figure:teaser}
\end{figure*}

Neural ranking models act as the core component of many search systems, often producing the final document scores. However, these scores are usually treated as transient information and only the relative orderings are preserved to produce a ranking. While this approach results in well-performing systems with respect to common retrieval metrics, such as nDCG and MAP, it ignores vital information that could be used by end users and downstream applications with real-world impacts, such as fair ranking~\citep{zerveas_mitigating_2022}, ranked list truncation~\citep{bahri_choppy_2020}, and query performance prediction~\citep{zhou_query_2007,shtok_predicting_2012,faggioli_query_2023}.

The common decision to discard the model scores comes from the fact that almost all neural ranking models, even probabilistic approaches, are trained to optimize relative orderings of documents as opposed to their absolute level of relevance. Aligning these ranking scores to a fixed scale is particularly difficult due to the nature of each query requiring different amounts of information to satisfy relevance, and it is much easier to determine if a document is more relevant than another rather than determining whether the information is sufficient for complex queries~\cite{khattab_colbert_2020,santhanam_colbertv2_2021,zhuang_rankt5_2023,laitz_inranker_2024}. 

% This property is why pairwise and listwise optimizations are so popular for ranking, as it is much easier to determine if a document is more relevant than another rather than determining whether the information is sufficient. 

This concept, that output scores should have meaningful real-world interpretations, is known as \emph{calibration}. Having this property ensures that a model’s predictions reliably reflects the ``true score.'' The most commonly recognized form of this is, where, for example, a classification model with a confidence of $p=0.47$ should be correct 47\% of the time. Building upon this idea of grounded scores, \textbf{scale calibration} extends this setting to values that do not have a direct probabilistic interpretation, such as document dwell time~\citep{smucker_time-based_2012} and multiple levels of relevance~\citep{yan_scale_2022,bai_regression_2023} which can exist beyond the [0,1] range.

Recent studies by~\citet{yan_scale_2022} and ~\citet{bai_regression_2023}, have highlighted a conflict between the objectives of the scale calibration and the ranking task, where optimizing for one may compromise the other. They propose new training objectives to improve scale calibration in learning-to-ranking (LTR) models, which are low in parameter count due to relying on these LTR datasets containing solely preprocessed features like length and query-term overlap as inputs. 
%This makes their findings difficult to directly apply to more effective yet larger neural ranking models. This situation raises a pivotal question for our research: What is the optimal strategy for scale calibration in neural ranking models\footnote{Here, ``neural ranking models'', or just ``neural rankers'', refers to neural models that produce ranking scores from textual query-document inputs, distinguishing them from deep feature-based LTR models that are significantly smaller.}?
% To address this question, we first identify several challenges associated with scale calibration of neural rankers, as opposed to traditional LTR models. Neural rankers often utilize advanced neural language models like BERT and T5~\citep{nogueira_passage_2019,nogueira_document_2019, nogueira_document_2020}, which makes them larger and more data-intensive compared to LTR models. A further challenge is the scarcity of training data suitable for scale calibration in neural rankers. Labeling data with nuanced levels of relevance requires substantial effort from skilled annotators or domain experts. Additionally, the complexity of the textual content in queries and documents introduces further difficulties. This complexity, coupled with the intricacy of raw information present in the texts, complicates the development of an effective mapping function from the input to the desired scale, especially with limited data.
Applying these methods to larger neural rankers over raw text remains a challenge.
%The main challenges in developing scale calibration methods for neural rankers compared to LTR lie in the disproportionate model size to data ratio. 
As opposed to LTR datasets, text-based collections with graded relevance judgements remain limited and orders of magnitude smaller. Further, the drastically higher parameter count of neural text rankers~\citep{craswell_trec_2021, nogueira_document_2019, nogueira_document_2020,zhuang_rankt5_2023,ma_fine-tuning_2023} make them even more data-intensive and reduces the efficacy of listwise loss functions due to hardware constraints and data sparsity.

In light of these challenges, we propose an alternative approach to improve scale calibration and score usability in these large text ranking models that does not rely on the loss functions of past works specific to the LTR settings. We instead ground the candidate query-document pairs via external knowledge, i.e., specifically leveraging the recent advances in generative and autoregressive large language models (LLMs) and their emerging capabilities for general language tasks with minimal supervision~\citep{wei_emergent_2022}.  
By producing natural language explanations (NLEs) of relevance with this additional knowledge, the ranking model will be able to leverage this information and produce better calibrated scores. As current LLMs have stability issues when generating text and can produce varying responses given the same input~\citep{ye_unreliability_2022,turpin_language_2023,chen_models_2023}, we hypothesize that the confidence and stability of a NLE generated from a zero-shot LLM {\em explaining} the relevance of a query-document pair correlates well with a calibrated ranking score. We validate this hypothesis by developing a method that aggregates the Monte Carlo (MC) generated NLEs of query-document relevance that serves as input into a traditional neural ranking model.
%aims to exploit this property by positing that the calibrated score of a query-document pair correlates with the confidence and uncertainty that a zero-shot LLM has in {\em explaining} their relevance. 
% We achieve this result by introducing an aggregated Monte Carlo (MC) generated NLE of query-document relevance. 
%In doing so, we capture essential information to determine relevance and use these as substitutes for the original query-document inputs. 
%We develop two types of natural language explanations: one that aligns with the language model's predictions and another that explicates both assumptions of relevance and irrelevance.

This work makes the following three contributions:
\begin{enumerate}
    \item Introduce MC-aggregate NLEs to improve calibration error and downstream usability of document rankings. 
    \item Significantly enhance the scale calibration of neural text rankers up to 25\% and downstream usability while maintaining or improving ranking effectiveness.
    \item Conduct a wide set of experiments on two benchmark text ranking datasets to validate performance and properties.
\end{enumerate}

%We conduct document ranking experiments on the TREC Deep Learning track~\citep{craswell_overview_2020} and the NTCIR-14 We Want Web-2 task~\citep{mao_overview_2019}, both containing meaningful multi-level relevance labels. Results demonstrate that LLM-generated NLEs significantly enhance the scale calibration of neural rankers, while maintaining or even boosting ranking performance in most scenarios.  The reduction of calibration error compared to previous approaches is up to $25\%$ on TREC and 16\% on NTCIR.

\section{Related Work}

\noindent \textbf{Scale Calibration of LTR Models.}  
Previous approaches to calibrating ranking models primarily involved post-processing methods~\citep{tagami_ctr_2013,chaudhuri_ranking_2017},  where a function maps the outputs for the frozen ranking model~\citep{platt_probabilistic_2000, zadrozny_transforming_2002}. Recent studies, such as those by \citet{yan_scale_2022} and \citet{bai_regression_2023}, have identified the issue that commonly used pairwise and listwise ranking objectives theoretically lead to uncalibrated scores and advocate for calibrated ranking objectives as a solution. While effective for the LTR setting that uses extracted features and small models, the limitations imposed by large text ranking models makes their effective application non-trivial.

\smallskip

\noindent \textbf{Uncertainty Calibration of Neural Ranking Models.} To achieve a notion of  calibration,  \citet{cohen_not_2021} and \citet{penha_calibration_2021} suggest leveraging a Bayesian interpretation of neural rankers to capture a distribution of possible scores for each document based on the internal uncertainty of the model. However, these approaches fail to ground the scores on any scale as such uncertainty information only makes sense when comparing documents, and is distinct from the type of calibration discussed in this paper.

\smallskip

\noindent \textbf{Natural Language Explanations.} NLEs are an emerging point of research in NLP and IR. These explanations can be presented directly to users to elucidate the rationale behind system decisions, such as in recommendation systems~\citep{li_generate_2020} and search results~\citep{yu_towards_2022}. More pertinently to this study, NLEs have been employed as intermediaries to enhance task efficacy~\citep{camburu_e-snli_2018,wei_chain--thought_2022,ferraretto_exaranker_2023}. \citet{ye_unreliability_2022} use the term overlap ratio between NLEs and their corresponding inputs as a feature to calibrate LLM predictions. \citet{wadhwa_using_2023} investigated explanation-based rescaling, converting coarse numeric scales into finer gradations. This technique leverages \emph{annotator-written} NLEs and scoring rubrics to prompt LLMs in generating rescaled outputs. Our research diverges by concentrating on NLEs {\em autonomously} generated by LLMs, which we leverage as inputs for neural rankers to facilitate scale calibration.

\smallskip

\noindent \textbf{Calibration of Large Language Models.} Various recent studies have focused on eliciting well-calibrated confidence scores from LLM responses. Some works treat LLMs as black boxes and use various prompting-based techniques to generate confidence scores~\cite{lin_teaching_2022, kadavath_language_2022}. Other works rely on training external models to associate confidence with LLM responses~\cite{kuhn_semantic_2022, kadavath_language_2022, mielke_reducing_2022}. \citet{tian_just_2023} conducted systematic evaluations over a wide range of methods to obtain confidence scores from LLMs and discovered that LLM-verbalized probabilities are better calibrated than conditional probabilities across several black-box RLHF-tuned LLMs. This framework was later extended to include hybrid methods (i.e., a mix of self-consistency and verbalized confidence) and evaluated the calibration performance on five different types of datasets~\citep{xiong_can_2023}. In Section~\ref{section:baselines}, we introduce two methods that take advantage of the calibration capabilities of LLMs to elicit scale calibrated ranking scores.

\section{Methodology} \label{section:method}

In this section, we formalize the definition of scale calibration of neural ranking models, following which we introduce the proposed method of leveraging natural language explanations (NLEs) for obtaining scale calibrated ranking scores. 

\subsection{Problem Statement and Motivation}
In ad-hoc ranking, we define a scoring function $\phi$ for a given query $q$ and its $n$ associated candidate documents $\{d^q\}_{1}^n$. This function, denoted as $\phi_{\Phi}(q, \{d^q\})$ %\xrightarrow{} \mathbb{R}^n$
produces a score for each query-document pair under the given retrieval model parameterized by $\Phi$. The ideal parameters of $\Phi$ are obtained by optimizing an empirical loss on a query-grouped training dataset $\mathcal{D} = \{(\{d^{q}\}, \{y^{q}\}) \, | \, q \in Q\}$. Here, $Q$ represents the set of training queries, and $\{y_q\}$ is the corresponding set of labels of documents $\{d^q\}$. The empirical loss is defined as:

\vspace{-6pt}
\begin{align}
% \small
\mathcal{L}(\Phi)=\frac{1}{|Q|}\sum_{q\in Q} l^{\text{rank}}(\{y^q\}, \phi_{\Phi}(q, \{d^q\})) \label{equation:training-loss}
\end{align}

\noindent where $l^{\text{rank}}$ is a ranking loss function for an individual query. In the case of neural ranking, the scoring function $\phi$ is defined by a backbone neural language model. For instance, using a pretrained BERT checkpoint as $\Phi$, concatenating query and each candidate document with a [\texttt{SEP}] token in between as the inputs, and leveraging cross entropy loss or pairwise cross entropy loss as $l^{\text{rank}}$ leads to the development of the widely known monoBERT or duoBERT models~\citep{nogueira_multi-stage_2019}, respectively. However, it has been observed that popular pairwise and listwise ranking losses are not scale calibrated due to their translation-invariant property~\citep{yan_scale_2022}\footnote{Although the cross entropy loss used in monoBERT is scale-calibrated, but it assumes only binary labels; which is something not observed in real-world datasets (e.g., multiple levels of relevance~\citep{craswell_overview_2020, mao_overview_2019}).}; adding a constant to all outputs of $\phi$ does not alter the loss value. %To partially overcome the issue of scale calibration in ranking loss, we incorporate the calibrated listwise softmax loss~\cite{yan_scale_2022}.

% We posit that merely employing a calibrated ranking loss for training is insufficient to fully address the scale calibration problem in neural rankers. 

Although calibrated ranking losses works well for LTR models~\citep{yan_scale_2022,bai_regression_2023},
%the high number of parameters in modern neural ranking models and the complexity of the input distribution make
the application of these approaches to neural text rankers is non-trivial. In LTR datasets with graded relevance judgements~\citep{chapelle_yahoo_2011,qin_introducing_2013,dato_fast_2017}, $|Q|$ is orders of magnitude larger than that of text based collections, which often only have less than $100$ queries. Further, $|\Phi|$ is much larger in neural text ranking models, limiting the use of  listwise loss functions due to GPU VRAM constraints. 
%
%In other words, the challenge is threefold: there is an imbalance in the parameter-to-data ratio, the neural ranking task itself is inherently more complicated than feature-based LTR, and the size of the model limits the use of loss functions due to hardware contraints.,
Addressing the scale calibration problem in neural rankers therefore requires a multifaceted approach that extends beyond the mere application of calibrated ranking loss.

\subsection{Scale Calibration via Natural Language Explanations}

We propose a novel two-step approach to obtain a scale-calibrated numerical score from the textual query and document: map the query-document to a NLE and then score this NLE using a standard neural ranking model.
This overall strategy is grounded in the recent success of LLMs to establish the relevance of query-document pairs~\citep{ferraretto_exaranker_2023} and the demonstrated efficacy of LLM-generated explanations in various reasoning tasks~\citep{wei_chain--thought_2022}.  
%In addition, zero-shot LLMs demonstrate exceptional ability to adapt to new tasks and domains with little to no efforts, making our approach more generalizable. 
We use an LLM, denoted as $g(\cdot)$, to generate NLEs $\{e^q\}$ for query-document pairs $\{d^q\}$, and then leverage a neural ranker over \emph{only} the NLEs to predict relevance scores. This process can be formally represented as a decomposition of $\phi$ into:
\vspace{-4pt}

\begin{align}
\phi_{\Phi}(q, \{d^q\}) &= f_{\Theta}(g_{\Psi}(q, \{d^q\})) \\
&= f_{\Theta}(\{e^q\}) \label{equation:ranker}
\end{align}

\noindent where $\Psi$ represents the parameters of the LLM, and $\Theta$ encapsulates the parameters of the neural ranker. The neural ranker $f(\cdot)$ in this setup is adapted to take the NLEs of the original inputs as its new inputs. Note that when using Eq.~\ref{equation:training-loss} to optimize parameters $\Phi=\{\Theta,\Psi \}$, we can optimize the parameters of the LLM and the neural ranker. While it is possible to perform full or partial fine-tuning on the LLM parameters $\Psi$, for simplicity and considering the limited amount of training data, we choose to freeze $\Psi$ and only optimize parameters of the neural ranker $\Theta$.

\subsection{Acquiring NLEs via LLM Prompting} 
\label{section:prompting}

Given the impact of prompting, we investigate two distinct methods for acquiring NLEs with varying characteristics.

\subsubsection{Literal Explanation} 

We present the query-document pair to the LLM and ask for both a relevance prediction (either ``relevant'' or ``non-relevant'') and an accompanying explanation. The exact prompt we use to generate such literal explanations is listed in Appendix~\ref{apx:prompt}.

A notable limitation of this approach is its susceptibility to inaccuracies stemming from the LLM's prediction errors. %For instance, if the LLM incorrectly labels a highly relevant query-document pair as ``non-relevant,'' the resulting NLE will not be useful to infer the true relevance score. 
To mitigate this, we employ a strategy involving Monte Carlo (MC) sampling of multiple NLEs for the same input and then forming a single meta NLE via an aggregation function $\texttt{AGGR}$ (detailed in Section~\ref{section:aggregation}):

\begin{align}
    e^q=\texttt{AGGR}(\{y_i \sim g_{\Psi}(y|q,d^q)\})
\end{align}

\noindent 
which aims to diminish the influence of erroneous predictions in the preference of the most likely generation. %While we find this method beneficial, we still observe a significant number of instances where the LLM consistently generates incorrect predictions in several MC samples.

\subsubsection{Conditional Explanation} \label{section:cond-exp}

To address instances where the LLM persistently errs in judging the relevance of an input, we also explore a different explanation generation strategy termed \emph{conditional explanation}. Motivated by the work of sampling multiple reasoning paths to enhance the self-consistency capabilities of LLMs~\citep{wang_self-consistency_2023} , this method involves prompting the LLM to generate rationales supporting both the relevance and non-relevance of a given query-document pair. The exact prompt we use to generate conditional explanations is also listed in Appendix~\ref{apx:prompt}.
Both conditional explanations are then used by the neural ranker $f(\cdot)$ to produce a scale calibrated ranking score (Eq.~\ref{equation:ranker}). %This approach allows for a more nuanced understanding and handling of relevance in scenarios where the LLM's dominant judgment may be skewed or overly confident.

\subsection{Aggregating Multiple NLEs} \label{section:aggregation}

As previously discussed, relying solely on the most probable output generated by the LLM, particularly in the literal explanation approach, could result in incorrect and overconfident explanations~\cite{ye_unreliability_2022}. To mitigate this issue, we propose sampling multiple generations from the LLMs. This technique has the advantage to introduce additional information, which may either support the initial judgment with varied reasoning or present contrasting viewpoints. We consider both outcomes to be advantageous: in the former scenario, it leads to a more robust and multi-faceted argument supporting a specific prediction; in the latter scenario, the generation of conflicting information by the LLM partially reveals its uncertainty in comprehending and assessing the relevance of the input query-document pair. We hypothesize that this uncertainty is indicative of predicting a more calibrated ranking score. 

Combining the information contained in the collection of Monte Carlo sampled NLEs produces 
%Combining these multi-faceted NLEs,In essence, any {\em novel} information found in less probable generations can be valuable and thus can be integrated into the construction of the NLE for the given input, culminating in what we term 
a {\bf meta NLE}. This meta NLE then serves as a more comprehensive and nuanced representation of the query-document relationship.
%
% The detailed algorithm and description regarding the NLE aggregation approach is described in Appendix~\ref{apx:aggr}.
%
\SetKw{Continue}{continue}
\SetKw{And}{and}
\SetKw{Break}{break}
\newcommand\mycommfont[1]{\footnotesize\ttfamily\textcolor{blue}{#1}}
\SetCommentSty{mycommfont}
\begin{algorithm}[t]
\caption{Novelty-based Aggregation}\label{algorithm:aggregation}
\SetKwHangingKw{KwInput}{Inputs:} % Set the Input
\SetKwHangingKw{KwOutput}{Output:}              % set the Output
\SetKwFunction{FMain}{AGGR}
\SetKwProg{Fn}{Function}{:}{}
\DontPrintSemicolon
\KwInput{$x$: input prompt; $\mathcal{E}$: sentence splitter; $\mathcal{S}$ \& $\lambda$: text similarity function \& threshold; $k_l$ \& $k_s$: sampling budgets; $g_{\Psi}(y|x)$: conditional output distribution defined by LLM.}
\KwOutput{Meta NLE $e$.}
\Fn{\FMain{$x,\mathcal{E},\mathcal{S},\lambda,k_l,k_s,g_{\Psi}$}}{
    $e \leftarrow \varnothing$ \\
    \For{$i \in 1,2,\cdots,k_l$}{
    $ y_i \leftarrow y \sim g_{\Psi}(y|x)$ 
    % \tcp*{Sample a new response from LLM}
    \\
    \For{$s \in \mathcal{E}(y_i)$}{
        \uIf{$e \neq \varnothing$ \And $\text{max}(\{\mathcal{S}(s,e_j);e_j \in e\}) > \lambda$}{
            \Continue
        } \Else{
            $e \leftarrow e \cup s$  
            % \tcp*{Add a novel sentence to meta NLE}
            \\
            \If{$|e| \ge k_s$}{ 
                \Return $e$  
                % \tcp*{Sampling budget reached}
                \\
            }
        }
    }
}
\Return $e$
}
\end{algorithm}

To create the meta NLE, we adopt an iterative approach (Algorithm~\ref{algorithm:aggregation}) to continually adds novel information. This process begins with an initially empty set of sentences (Line 4). If a sentence from the newly generated explanation (Line 6-7) introduces novel information – as determined by its maximum text similarity to the existing sentences in the set being at or below a predefined threshold – then this sentence is added to the set (Line 11). This sampling of new explanations continues until we either reach the predefined maximum number of sampling iterations (Line 5) or fulfill the limit for the number of sentences in the meta NLE set (Line 12).  This approach is able to latently encode the uncertainty of the LLM, as a high variance set of sampled NLEs will inevitably produce a longer meta NLE.

\section{Experiments}

In this section, we assess our proposed methods using two real-world, extensively used document ranking datasets to evaluate the following research questions (RQ):

\begin{enumerate}[font={\bfseries},label={RQ\arabic*:}]
\item How do LLM-generated natural language explanations impact the calibration and ranking performance of neural rankers?
\item Is there a consistent improvement across different training objectives when using these explanations?
\item Does the MC sampled aggregate-NLEs from the LLM yield empirical improvements?
\item Does the proposed methodology produce meaningful scores in a real-world setting?
\end{enumerate}

%This experimental framework is designed to not only quantify the effectiveness of LLM-generated NLEs in enhancing neural rankers but also to explore the broader applicability and potential constraints of our methods.

\subsection{Data}

We employ two widely acknowledged datasets in information retrieval research: the TREC Deep Learning Track (\textbf{TREC})~~\citep{craswell_trec_2021}, covering the period from 2019 to 2022, and the NTCIR-14 WWW-2 (\textbf{NTCIR})~\citep{mao_overview_2019}. 

The choice of these datasets is motivated by their comprehensive multi-level relevance judgments provided by human annotators and the ample volume of labeled documents for each query. Specifically, TREC uses passages from the MS MARCO collection~\citep{bajaj_ms_2018}, while NTCIR employs web pages from ClueWeb12 Category-B.\footnote{\url{https://lemurproject.org/clueweb12/}} For TREC, we partition the queries from the years 2019 and 2020 for training, use the 2021 queries for validation, and the 2022 queries for testing. For NTCIR, the queries are divided into training, validation, and testing sets in a 6:2:2 ratio. Table~\ref{table:datasets} shows provides the statistics for these collections.

\begin{table}[t]
\centering
\begin{adjustbox}{width=\linewidth}
\begin{tabular}{l|cc}
\hline
\textbf{Metric} & \textbf{TREC} & \textbf{NTCIR} \\ \hline
\# Q (train/val/test) & 97/53/67 & 48/16/16 \\ \hline
Avg. \# docs per query & 282.7 & 345.3 \\ \hline
Levels of relevance & 4 & 5 \\ \hline
Label dist. (low to high) & 58/22/14/6 & 48/23/17/8/3 \\ \hline
Avg. query length & 8.0 & 22.0 \\ \hline
Avg. doc. length & 70.9 & 493.2 \\ \hline
\end{tabular}
\end{adjustbox}
\caption{Statistics of the TREC and NTCIR datasets. 
% The lengths of queries and documents are quantified using BERT tokenization. For the NTCIR dataset, documents sourced from ClueWeb have undergone preprocessing to retain only the initial 512 tokens.
} \label{table:datasets}
\end{table}

\subsection{Metrics} \label{section:metrics}

Our goal is to devise methods that not only enhance the calibration of neural rankers but also maintain their ranking effectiveness. Accordingly, we evaluate both calibration and ranking aspects. 

\smallskip
\noindent
\textbf{Ranking Metrics:} We employ the normalized discounted cumulative gain ({\bf nDCG}), which accounts for multiple levels of relevance judgment. We report \textbf{nDCG} and \textbf{nDCG@10}.

\smallskip
\noindent
\textbf{Calibration Metrics:} For calibration effectiveness, mean square error ({\bf MSE}), as well as the expected calibration error ({\bf ECE})~\citep{guo_calibration_2017}, a metric initially developed for \emph{classification} calibration, are commonly used in ranking literature~\cite{yan_scale_2022,bai_regression_2023,penha_calibration_2021}. ECE measures the discrepancy between a model's predictions and gold labels in a binning approach via
\begin{equation}
% \small
 ECE = \sum_{m=1}^{M} \frac{|B_m|}{n} \left|acc(B_m) - conf(B_m)\right|,
\end{equation}
 where $M$ is the number of bins $B$. However, given the highly skewed and unbalanced label distribution in our test set, as indicated in Table~\ref{table:datasets}, MSE and ECE  are biased. Opting for scores that are near the most frequent label results in lower errors while having no real world usability. This bias undermines these metrics' effectiveness in the ranking context, and is further discussed in Appendix~\ref{apx:ece}. To achieve a metric that more closely aligns with real world usability for the search task, we propose the adoption of a class-balanced version of ECE, called {\bf CB-ECE}, which allocates equal significance to all candidate scale values.  The CB-ECE is computed by initially calculating the ECE for each individual class, and subsequently taking the mean,
\begin{equation}
% \small
\text{CB-ECE} =\frac{1}{C} \sum_{c=1}^{C} ECE_c,
\end{equation}
with $ECE_c$ as the error for each class, $c$.

\smallskip
\noindent
\textbf{Query Performance Prediction (QPP):} As the motivation of this work comes from the desire of usable scores, we also evaluate the methods using the QPP task. We evaluate the effectiveness of each approach by using \textbf{Pearson's $\rho$} and \textbf{Kendall's $\tau$ }between the predicted and actual query performance.

\begin{table*}[t]
\begin{adjustbox}{width=\linewidth,center}
\begin{tabular}{cccccc|ccccc}
\toprule
  \multicolumn{1}{c}{Collection} &
  \multicolumn{5}{c}{TREC} &
  \multicolumn{5}{c}{NTCIR} \\
\midrule
  \multicolumn{1}{c}{\multirow{2}{*}{Metric}} &
  \multicolumn{2}{c}{Ranking} &
  \multicolumn{3}{c}{Calibration} &
  \multicolumn{2}{c}{Ranking} &
  \multicolumn{3}{c}{Calibration} \\
           & nDCG($\uparrow$)  & nDCG@10($\uparrow$) & CB-ECE($\downarrow$) & ECE($\downarrow$)   & MSE($\downarrow$)   & nDCG($\uparrow$)  & nDCG@10($\uparrow$) & CB-ECE($\downarrow$) & ECE($\downarrow$)   & MSE($\downarrow$)   \\
\midrule
{\bf NC}          & 0.799 & 0.494   & 1.205  & 0.320 & 0.773 & 0.735 & 0.337   & 1.757  & 0.799 & 1.824 \\
\midrule
{\bf PC}         & 0.799	& 0.494	  & 1.141  & 0.125 & 0.684 & 0.735 & 0.337   & 1.624  & 0.457 & 1.462 \\
\midrule
{\bf FC} (monoBERT)  & 0.776 & 0.422   & 1.093  & 0.221 & 0.721 & 0.696 & 0.268   & 1.843  & 0.709 & 1.874 \\
{\bf FC} (BERT)      & 0.738 & 0.327   & 1.253  & 0.266 & 0.785 & 0.727 & 0.285   & 1.756  & 0.546 & 1.416 \\
\midrule
{\bf PR}                        & 0.786 & 0.457   & 1.000  & 1.246 & 2.137 & 0.728     & 0.328       & {\bf 1.294}      & 1.194     &  2.773    \\
\midrule
{\bf PL}   & 0.790 & 0.473   & 1.165  & 0.145 & 0.673 & 0.736 & {\bf 0.364}   & 1.677  & 0.472 & 1.540 \\
\midrule
 {\bf NLE} (literal)  & 0.815$^\dagger$ & 0.529$^\dagger$   & 0.996  & {\bf 0.067}$^\dagger$ & {\bf 0.602}$^\dagger$ & {\bf 0.742} & 0.340   & 1.534  & 0.355$^\dagger$ & 1.330$\dagger$ \\
 {\bf NLE} (conditional) & {\bf 0.822}$^\dagger$ & {\bf 0.534}$^\dagger$ & {\bf 0.862}$^\dagger$ & 0.428 & 0.832 & 0.720 & 0.322 & 1.405 & {\bf 0.257}$^\dagger$ & {\bf 1.290}$^\dagger$ \\
\bottomrule
\end{tabular}
\end{adjustbox}
\caption{Ranking and scale calibration performance on TREC and NTCIR. Note that lower is better with calibration metrics (CB-ECE, ECE and MSE). Statistically significant improvements (t-tests with Bonferroni correction at the 95\% confidence level) over all baselines are marked with $^\dagger{}$.  \label{table:main}
}
\end{table*}

\subsection{Baselines} \label{section:baselines}

Due to the scarcity of relevant work in scale calibrated search, we have developed several methods to establish meaningful baselines in addition to the past relevant work of~\citet{yan_scale_2022}. For consistency, we use the same neural ranking architecture, when applicable, for all approaches. Additional details are in Appendix~\ref{apx:baselines}.

\smallskip

\noindent \textbf{No calibration (NC)}. We use monoBERT \citep{nogueira_passage_2019}, a BERT-based ranker that has been previously finetuned with the MS MARCO passage ranking training set.

\smallskip

\noindent \textbf{Post-hoc calibration (PC)}. We apply Platt scaling~\citep{platt_probabilistic_2000} adapted for regression calibration to the uncalibrated monoBERT. It is important to note that the parameters of the ranker remain fixed under this approach.

\smallskip

\noindent \textbf{Full calibration (FC)}. We use the training data to finetune a BERT-based ranker using calibrated softmax loss~\citep{yan_scale_2022}, which represents a straight adoption of the approach from LTR to neural ranking. As discussed, due to substantially larger size of the neural ranker and reduced training data, we are forced to only use a small fraction of negatives for each query compared to under the LTR setting. %The BERT model here can either be the monoBERT mentioned earlier, or a BERT model that is not finetuned for ranking\footnote{\url{https://huggingface.co/bert-base-uncased}}.

\smallskip 

\noindent \textbf{LLM prompting with rubrics (PR)}. Inspired by studies on using LLMs for rendering relevance judgments~\citep{zhuang_beyond_2023, thomas_large_2023, faggioli_perspectives_2023} and rescaling~\citep{wadhwa_using_2023}, we explore leveraging the scoring rubrics as a contextual guide for LLM prompting. We present each query-document pair, along with the rubrics from each dataset that were initially used to instruct human annotators, to the LLM and take the scalar predictions.

\smallskip

\noindent \textbf{Post-hoc calibrating LLM predictions (PL)}. Based on prior studies that LLM confidence scores are well-calibrated~\citep{kadavath_language_2022,tian_just_2023}, we use the LLM as a zero-shot binary classifier and take the average predictions over 20 samples as the confidence score. We then use Platt scaling~\citep{platt_probabilistic_2000} to map these scores to the target scale.

\smallskip 

\noindent \textbf{Full calibration with NLEs (NLE)}.  We first acquire meta NLEs for candidate query-document pairs and then fine-tune a BERT model\footnote{We experimented with using monoBERT to initialize this ranker but found it to perform worse than general-purpose BERT weights.} using the NLEs as inputs with the calibrated softmax loss with the same restrictions as FC. This forms a direct comparison with FC, as they use the same initialization weights, labels and training objectives, but different inputs.

\begin{figure*}[h]
    \centering
    \includegraphics[width=0.98\linewidth]{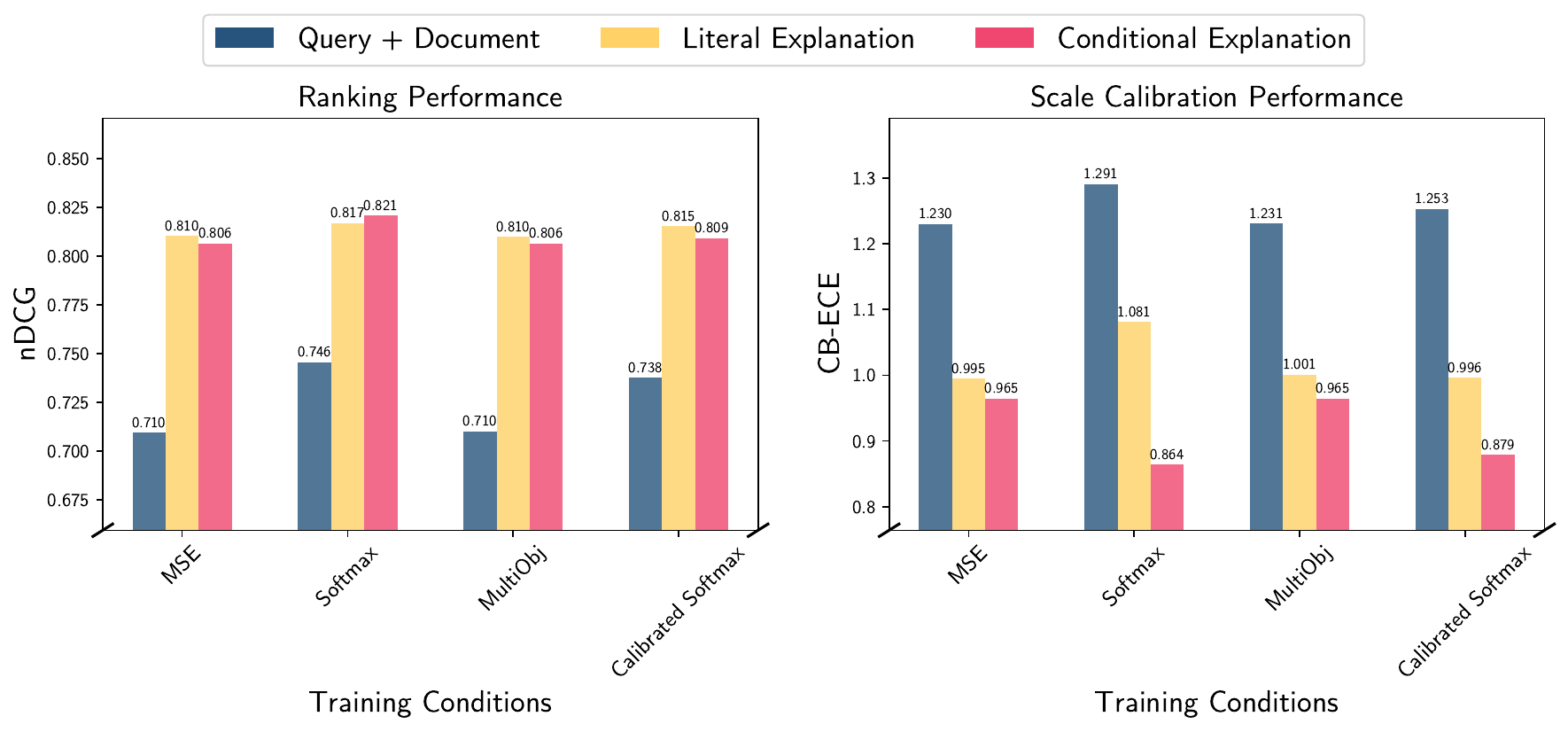}
    \caption{Ranking and scale calibration performance on TREC of full calibration of BERT, taking query + document inputs (FC BERT) and our proposed explanations, using four different optimization objectives. NLE-based approaches consistently yield better ranking (left) and calibration (right) performance.}
    \label{figure:training-losses}
\end{figure*} 

\subsection{Downstream Performance: QPP}

The ultimate goal of a calibrated model is interpretable scores by end users and algorithms. Therefore we evaluate the scores produced from each approach using the query performance prediction (QPP) task. The goal of QPP, as discussed by~\citet{arabzadeh_query_2024}, is to estimate the performance of individual queries such that hard-to-satisfy queries can be better addressed by search systems. We use two popular post-hoc QPP methods—Weighted Information Gain (WIG)~\citep{zhou_query_2007} and Normalized Query Commitment (NQC)~\citep{shtok_predicting_2012}—to explore whether the proposed NLE based approaches yield more usable scores. Evaluations are conducted on the test set of TREC, using the top-10 documents to calculate WIG and NQC.

\subsection{Reproducibility}

We leverage the open-source LlaMA2-13B-Chat model~\citep{touvron_llama_2023} as the LLM and the BERT model~\citep{devlin_bert_2019} as the neural ranker. Implementation details are described in Appendix~\ref{apx:impl}. The code repository can be found at \url{https://github.com/PxYu/LLM-NLE-EMNLP}.
% The anonymized code repository can be found at \url{https://anonymous.4open.science/r/LLM-NLE-Ranking}.

\subsection{Results and Analysis}

\textbf{RQ1: How do LLM-generated natural language explanations impact the calibration and ranking performance of neural rankers?} We find in  Table~\ref{table:main} that NLE-based methods yield statistically significant improvements for both aspects. For scale calibration, NLEs exhibit lower CB-ECE compared to both post-hoc calibration (PC) and full calibration (FC) across both datasets, with literal NLEs performing better on TREC while conditional NLEs perform slightly better on NTCIR. Regarding ranking, the proposed approach significantly surpasses NC and FC baselines, which shares the same fine-tuning data and initial weights, with improvements of up to 11.4\% in nDCG and 63.3\% in nDCG@10 in the TREC dataset. This indicates that LLM-generated NLEs provide valuable insights for document differentiation beyond the scope of calibration. %Examining performance on the NTCIR dataset, the conditional explanation approach slightly underperforms compared to monoBERT in terms of ranking, whereas the literal explanation approach still showcases significant performance improvements over monoBERT, which is most likely attributed to the aggregation method. % In summary, RQ1 can be affirmatively answered: LLM-generated NLEs significantly enhance the scale calibration of neural rankers, often maintaining or even boosting ranking performance in most scenarios.

\smallskip

\noindent \textbf{RQ2: Is there a consistent improvement across different training objectives when using these explanations?} \citet{yan_scale_2022} and \citet{bai_regression_2023} advocate for addressing the scale calibration of LTR models through calibrated loss functions. As a significant motivator for this work are the challenges of applying LTR approaches to larger text models, we investigate the efficacy of naively applying these objectives to the neural ranking task.
To this end, we experiment with four fine-tuning objectives commonly used in this setting: MSE as a pointwise regression loss, uncalibrated listwise softmax (Softmax), a multi-objective combination of MSE and Softmax (MultiObj) from \citet{bai_regression_2023}, and the calibrated listwise softmax (Calibrated Softmax) from~\citet{yan_scale_2022}. The FC baseline is also trained with each of these objectives. As seen in Fig.~\ref{figure:training-losses}, the NLE-based approaches consistently surpass the performance of neural models that process raw text queries and documents across all four optimization methods. Further, while these objectives do produce different results, the majority of the improved calibration performance comes from the NLEs themselves and is relatively agnostic of the loss function. 
%This outcome underscores the effectiveness, versatility and robustness of our explanation acquisition methods for calibrating neural rankers, positioning them as significant contributions distinct from those focused on calibrated loss functions. 
% Furthermore, the consistent improvement across different training objectives highlights the versatility and robustness of our proposed approaches.

% \subsection{Impact of NLE Selection}

% \noindent \textbf{RQ3 and RQ4}.
\smallskip
\noindent
\textbf{RQ3: Does the MC sampled aggregate-NLEs from the LLM yield empirical improvements?} To further the understanding of the dynamics between LLM-generated NLEs and their impact on scale calibration and ranking, we perform additional experiments on the TREC dataset.  
% discussed in Section~\ref{section:aggregation},
We evaluate the decision to use an aggregate NLE rather than the most probable NLE (discussed in Section~\ref{section:aggregation}) in Table~\ref{table:ablation}. We include an oracle baseline, where we repeatedly sample a NLE until we achieve one that aligns with the true relevance label. If this is not achieved within 20 samples, we default to the most confident NLE even if it's incorrect. It can be observed from the results that, across both literal and conditional NLE settings, using the proposed aggregate MC method achieves better performance both w.r.t. ranking as well as calibration.
%is employed as input for the neural ranker. Our investigation also extends to understanding the influence of the LLM’s inherent limitations, particularly in its judgment alignment with human annotators~\citep{faggioli_perspectives_2023,thomas_large_2023}, on the performance of downstream NLE-based neural rankers. In our literal explanation approach, we keep sampling responses from the LLM, containing both prediction and explanation, until we obtain a response that aligns with the binary relevance judgment provided by annotators. Specifically, a match occurs when the LLM predicts ``relevant'' for labels 1, 2, or 3, and ``non-relevant'' for label 0. If alignment is not achieved within 20 samples, we default to using the most probable, albeit incorrect, explanation. This experimental condition is denoted as the \textbf{oracle} setting. The results of these experimental setups are presented in Table~\ref{table:ablation}.

\begin{table}[t]
\begin{adjustbox}{width=0.9\linewidth,center}
\begin{tabular}{llcc}
\toprule
Explanation                        &  Selection             & nDCG($\uparrow$)  & CB-ECE($\downarrow$) \\ \midrule
\multirow{3}{*}{Literal}  & Most Probable & 0.789 & 1.093      \\
                             & Aggregate MC  & {\bf 0.815}$^\dagger$ & {\bf 0.996}$^\dagger$  \\
                             & \textcolor{gray}{Oracle}        & \textcolor{gray}{0.883} & \textcolor{gray}{0.801}  \\
\midrule
\multirow{2}{*}{Conditional} & Most Probable & 0.797 & 0.895  \\
                             & Aggregate MC  & {\bf 0.822}$^\dagger$ & {\bf 0.862}$^\dagger$ \\
\bottomrule
\end{tabular}
\end{adjustbox}
\caption{The effect of different types of natural language explanations and selection strategies on the ranking and scale calibration performance of neural rankers.} \label{table:ablation}
\end{table}

%\noindent {\bf RQ3: Can sampling multiple instances of NLEs from the LLM yield empirical improvements?}. The results clearly demonstrate that the aggregate MC method substantially surpasses the use of only the most probable explanation in both ranking and scale calibration, across both literal and conditional explanation setups. This finding supports our hypothesis regarding the benefits of MC sampling multiple explanations. It also underscores the effectiveness of our aggregation strategy, affirming its utility in this context. 

%\noindent {\bf RQ4: What inherent limitations are posed by the LLM in our approaches?}. A notable performance disparity is observed between our current aggregate MC approach and the oracle explanation setting. This gap highlights a significant potential for improvement in NLE-based neural rankers, contingent upon better alignment of LLM judgments with human annotators on a \emph{binary} scale. Specifically, enhancements in this alignment could be achieved through methods such as fine-tuning, prompt engineering, or employing LLMs with more advanced reasoning capabilities. We leave the exploration of this direction for future work.

% \subsection{Downstream Application: Query Performance Prediction}

\smallskip

\noindent
\textbf{RQ4: Does the proposed methodology produce meaningful scores in a real-world setting?}
%As an example, we consider the post-hoc query performance prediction (QPP) task, where document ranking scores are leveraged to predict the difficulties of search queries. The goal of QPP, as discussed by~\citet{arabzadeh_query_2024}, is to estimate the performance of individual queries such that hard-to-satisfy queries can be better addressed. We use two popular post-hoc QPP methods—Weighted Information Gain (WIG)~\citep{zhou_query_2007} and Normalized Query Commitment (NQC)~\citep{shtok_predicting_2012}—to explore whether using more scale-calibrated ranking scores would enhance the accuracy of QPP models. To evaluate the accuracy of QPP, we use Pearson’s $\rho$ and Kendall’s $\tau$ correlations between predicted and actual query performances. Evaluations are conducted on the test set of TREC, using the top-10 documents to calculate WIG and NQC, and nDCG@10 as the measure of actual query performance.
The results from using ranking scores from different calibration approaches for QPP are reported in Table~\ref{table:qpp}, and we are able to definitively show that the scores produced via NLEs result in substantially more usable scores. Interestingly, directly adapting the calibrated ranking loss of \citet{yan_scale_2022} from LTR to neural ranking (FC) leads to much worse QPP performance compared to no calibration (NC). PC, PR, and PL, though improving calibration metrics, do not consistently improve QPP either. Comparing between the literal and conditional NLE methods, there is a large gap in performance suggesting that including the two extreme perspectives of the conditional NLEs play a critical role. We highlight that while MSE, ECE and CB-ECE are often used in past works for ranking calibration, it leaves out critical information for the ranking task. We hope to investigate more effective metrics in future work.

%improves QPP in most cases, while the conditional NLE approach demonstrates superior performance across all QPP methods and metrics. Therefore, we conclude that better scale calibration does not necessarily lead to better downstream post-hoc QPP performance, yet the NLE-based approaches proposed here are effective.

\begin{table}[t]
\begin{adjustbox}{width=0.75\linewidth,center}
\begin{tabular}{@{}c|cc|cc@{}}
\toprule
                  & \multicolumn{2}{c|}{WIG} & \multicolumn{2}{c}{NQC} \\ \midrule
Metric                  & P-$\rho$   & K-$\tau$   & P-$\rho$    & K-$\tau$   \\ \midrule
NC                &      0.176      &      0.091      &     0.003       &      -0.083      \\
PC                &    0.198        &      0.113      &      -0.173      &      -0.123      \\
FC                &     -0.001       &     -0.012       &   -0.385         &     -0.242       \\
PR                &     0.052       &    0.003        &    -0.002        &     -0.025       \\
PL                &    0.019       &    0.057        &     -0.262       &     -0.101       \\
NLE (lit.)     &     0.063       &     0.104       &      0.075      &     0.048       \\
NLE (con.) &   \bf 0.268         &     \bf 0.165       &     \bf 0.183      &      \bf 0.146     \\ \bottomrule
\end{tabular}
\end{adjustbox}
\caption{The effect of scale calibration on post-hoc QPP.} \label{table:qpp}
\end{table}

\section{Conclusions}

This study tackled the challenge of scale calibration for neural rankers.
% , aiming to align ranking scores with meaningful real-world measures. 
Our method leverages zero-shot LLMs' inherent understanding of textual data through NLEs to enhance scale calibration while maintaining or improving ranking performance. We showed that our approach surpasses established baselines in ranking and calibration metrics, demonstrates consistency across various training objectives, and is effective for downstream post-hoc QPP applications.

% Despite the effectiveness of using zero-shot LLMs, there is potential for improvement through more sophisticated strategies like few-shot prompting~\citep{wei_emergent_nodate}, instruction tuning~\citep{zhu_inters_2024}, and the use of LLMs with advanced reasoning capabilities. Future research could also focus on increasing the efficiency of NLE generation through techniques like distillation \citep{gu_knowledge_2023,shridhar_distilling_2023}. 
% Furthermore, enhancing the reliability of explanations \citep{ye_unreliability_2022} represents another promising avenue for developing better calibrated rankers.

\clearpage

\section{Limitations}

This work prototypes the idea of leveraging LLM-generated NLEs of query-document pairs to improve the scale calibration of neural ranking models. Like many recent works that incorporate LLMs as an integral part of the document ranking process~\citep{ma_zero-shot_2023,qin_large_2023,zhuang_setwise_2024}, our approach is not yet efficient enough to be directly applied to an online ranking system with high latency requirements. Having validated the feasibility of NLEs for calibration, future work could incorporate techniques such as distillation to improve run-time efficiency.

Another limitation stems from the unreliability and inconsistency of NLEs from LLMs~\citep{ye_unreliability_2022}. Although we paradoxically leverage the inconsistency of NLEs to further improve the scale calibration of neural ranking models, the potential effects of addressing unreliability or inconsistency on the calibration task remain unclear. Moreover, while we aim for the scale-calibrated ranking \emph{scores} to be useful for explainability purposes as they correspond to certain criteria, the NLEs themselves are not meant to be, nor should they be directly displayed to users for interpretability purposes in the current form.

A notable performance disparity is observed between our current aggregate MC approach and the oracle explanation setting in Table~\ref{table:ablation}. This gap highlights a significant potential for improvement in NLE-based neural rankers contingent upon better alignment of LLM judgments with human annotators on a \emph{binary} scale. Although we experimented with various prompts for generating NLEs in some pilot studies, this research is designed to explore the feasibility of LLM-generated NLEs for a specific important task. Consequently, various directions that could potentially improve the LLMs in terms of accuracy of predictions and quality of explanations—including but not limited to few-shot prompting, prompt tuning and optimization, LLM fine-tuning, and leveraging better (even closed-source) LLMs—are beyond the interest and scope of this study.

% Specifically, enhancements in this alignment could be achieved through methods such as fine-tuning, prompt engineering, or employing LLMs with more advanced reasoning capabilities.

% Finally, although we experimented with various prompts for generating NLEs in some pilot studies, this research is a prototype designed to explore the feasibility of LLM-generated NLEs for a specific important task. Consequently, various directions that could potentially improve the quality of explanations—including but not limited to few-shot prompting, prompt tuning and optimization, LLM fine-tuning, and leveraging better and even closed-source LLMs—are beyond the interest and scope of this study.

\section{Potential Risks} 
Although calibrated ranking scores can be leveraged in certain ways to mitigate bias for fairer document ranking~\citep{heuss_predictive_2023}, the use of NLEs acts as an information transformation, filtering parts of the candidate document while adding the LLM's own biases. Certain topics, demographics, and perspectives could be impacted and consequently change the proposed document rankings. While traditional \textit{ad-hoc} web search is transitioning to retrieval augmented generation, the consequence remains the same. These systems act as information brokers for the general public, and can easily introduce or reinforce biases into the population based on the retrieved information. With this in mind, we elected to use an off-the-shelf text ranking paradigm as the final scorer.  This facilitates the drop in use of the many techniques already developed to improve fairness in ranking~\citep{zehlike_fairness_2023-1,zehlike_fairness_2023}.

\clearpage

% Bibliography entries for the entire Anthology, followed by custom entries
%\bibliography{anthology,custom}
% Custom bibliography entries only
\bibliography{zotero}

\appendix
\section{Prompts Used in Experiments}
\label{apx:prompt}

The format of the prompt\footnote{We experimented with different prompts, but stick to this one for clear performance gains and ease of formatting.} we employ for literal explanation is as follows:

\smallskip

\noindent \textcolor{gray}{``For the following query and document, judge whether they are relevant or non-relevant, and provide an explanation. Output `Relevant' or `Nonrelevant'. Do not repeat the content of the query or the document. Query: \{query\} Document: \{document\} Output:''}

\smallskip

This method parallels the prompt used by~\citet{ferraretto_exaranker_2023} to generate explanations for query-document relevance for training generative rankers. However, our approach differs due to its simplicity and broader generalizability across datasets, utilizing zero-shot prompting in contrast to their use of a fixed set of 7 examples as demonstrations for few-shot prompting.

The prompt we use for the conditional explanation approach is as follows:

\smallskip

\noindent \textcolor{gray}{``For the following query and document, explain why they are \{relevant/nonrelevant\}. Query: \{query\} Document: \{document\} Output:''}

\section{Details of Competing Methods} \label{apx:baselines}

Here we provide more details and explanations for competing methods evaluated in the experiments.

\smallskip

\noindent
\textbf{No calibration (NC)}.
Our approach here employs a BERT-based ranker that has been previously finetuned using the MS MARCO dataset~\citep{bajaj_ms_2018}. Despite its strong performance in ranking tasks on TREC (in-domain) and NTCIR (out-of-domain), this ranker lacks scale calibration. This is attributed to the binary relevance labels in MS MARCO, which differ from the multi-level relevance judgments used in TREC and NTCIR.

\smallskip

\noindent
\textbf{Post-hoc calibration (PC)}.
In this method, we adjust the output scores of the finetuned rankers using a learnable function. Following the approach of~\citet{yan_scale_2022}, we apply Platt scaling~\citep{platt_probabilistic_2000}, adapted for regression calibration. Given the output ranking scores of the model $\mathbf{s}$, the calibrated scores are computed as $\mathbf{s'}=\exp{(w\mathbf{s}+b)} / 2$, where $w$ and $b$ are learnable parameters. It is important to note that under this method, the parameters of the finetuned rankers remain fixed; only the parameters $w$ and $b$ are optimized using the scale calibration data. This approach maintains the original ranking performance of the ranker, provided the final value of $w$ is positive.

\smallskip

\noindent
\textbf{Full calibration (FC)}. This strategy involves directly fine-tuning a BERT-based ranker with ``query-[\texttt{SEP}]-document'' style inputs on scale calibration data. Utilizing the monoBERT checkpoint allows for a direct comparison with PC methods - they use the same initial checkpoint and training data, but optimize different parameters. Starting with a general-purpose BERT checkpoint sets up a direct comparison with our NLE-based calibration approaches - they share the same initial weights and training data, but the format of their input data differs significantly.

\smallskip

\noindent
\textbf{LLM prompting with rubrics (PR)}. The underlying concept is that the LLM should be able to comprehend the given rubrics and automatically produced calibrated scores that align with the rubrics, thereby removing the need for any post-hoc calibration steps. It is important to note that in this method, the LLM itself essentially functions as the ranker, without the integration of an additional neural ranking model. For the TREC dataset, we adopt the prompt method used by~\citet{thomas_large_2023}. Conversely, for NTCIR, we adapt the approach to incorporate the specific rubric outlined in the task description of NTCIR-14 WWW-2~\citep{mao_overview_2019}. It is noteworthy that the NTCIR rubric is defined based on scores provided by two annotators (e.g., ``relevance=3: One annotator rated as highly relevant, one as relevant''), focusing on the quantitative synthesis of annotations. In contrast, the TREC rubric is more qualitatively oriented, emphasizing the explanation of query-document relationships.

\smallskip
\noindent
\textbf{Post-hoc calibrating LLM predictions (PL)}.
In this method, we use the LLM as a zero-shot classifier to determine whether a given query and document pair is relevant (denoted as 1) or not (denoted as 0). To mitigate the bias inherent in the most probable generation and to minimize instances of tied scores, which complicate the derivation of rankings, we sample the LLM's responses 20 times for each input and calculate the average of these scores. Subsequently, we employ Platt scaling to these averaged scores and refine the parameters using the training set. The prompt we use is similar to that of~\citet{zhuang_beyond_2023}, with one significant modification: we instruct the LLM to output either ``relevant'' or ``nonrelevant'' in lieu of ``yes'' or ``no''. This alteration stems from our observation that the LLM exhibits a strong prior towards generating affirmative responses such as ``Yes, I can help you with this request...'', which could potentially skew the predictions. By specifying the terms ``relevant'' and ``nonrelevant'', we aim to reduce this bias and achieve more accurate relevance predictions.

\smallskip
\noindent
\textbf{Full calibration with NLEs (NLE)}. Building upon our methods for generating and aggregating natural language explanations (NLEs) for query-document pairs, as discussed in Section~\ref{section:method}, we proceed to finetune a BERT model (not finetuned for retrieval) to process meta NLEs and yield scale-calibrated ranking scores. In the scenario of the conditional explanation approach (Section~\ref{section:cond-exp}), where each input is represented with two meta NLEs, one for relevance and one for non-relevance, our method involves an additional processing step. Specifically, we concatenate the \texttt{[CLS]} hidden states obtained from encoding both NLEs. This concatenated representation then feeds into an additional linear layer, which is responsible for transforming these combined hidden states into a final ranking score. This approach allows the model to integrate insights from both relevance perspectives.

\section{Class-balanced ECE and the limitations of ECE} \label{apx:ece}

We discuss in Section~\ref{section:metrics} the limitations inherent in using the expected calibration error (ECE) \citep{guo_calibration_2017} for scale calibration data, especially when faced with imbalanced label distributions. To counter this challenge, we propose a straightforward modification: the adoption of class-balanced ECE (CB-ECE). This section uses model outputs on TREC as illustrative examples to provide detailed insights and empirical support for our approach.

\begin{figure}[htbp]
  \centering
  \begin{subfigure}[b]{0.5\linewidth}
    \includegraphics[width=\linewidth]{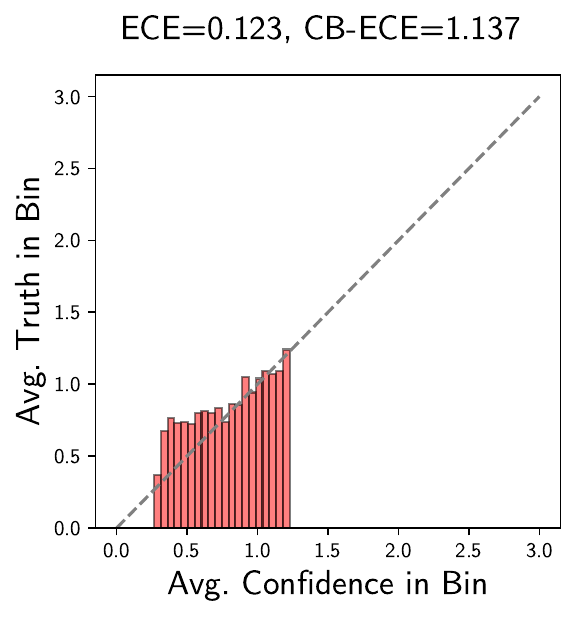}
    % \caption{First Subfigure}
    \label{fig:sub1}
  \end{subfigure}%
  \begin{subfigure}[b]{0.5\linewidth}
    \includegraphics[width=\linewidth]{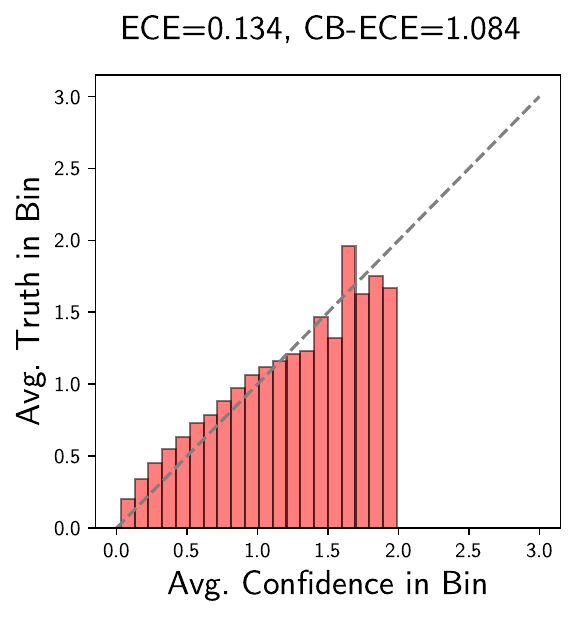}
    % \caption{Second Subfigure}
    \label{fig:sub2}
  \end{subfigure}
  \caption{Reliability diagrams for two models on TREC: The left diagram shows a model with ranking scores densely concentrated on the lower part of the scale, which exhibits better ECE performance due to ECE's failure to account for prediction coverage across the target scale. On the right, the CB-ECE penalizes this undesirable behavior, indicating that the model providing better coverage across the scale is more effectively calibrated.}
  \label{figure:reliability-diagram}
\end{figure}

We employ reliability diagrams~\citep{murphy_reliability_1977}, a widely-used technique in calibration, to visualize calibration performance. In ECE calculations, samples are sorted by their predicted values and divided into several buckets. The calibration error for each bucket is determined by the absolute difference between the mean labels and mean predictions of samples within that bucket, further \emph{weighted by} the number of samples falling in the bucket. Reliability diagrams concretely visualize this concept, with each bucket represented by a bar. The mean predictions are indicated on the x-axis, while the mean labels (i.e., scale) are represented by the height of the bars. In an ideal scenario, a perfectly calibrated model would yield a reliability diagram aligning with the diagonal line, indicating that mean predictions correspond accurately to the labels for every bucket. However, ECE overlooks a critical aspect of scale calibration: \textbf{coverage}. This refers to the model's ability to adequately span the target scale range. Figure~\ref{figure:reliability-diagram} illustrates this point with two diagrams. The left diagram shows a model whose outputs are narrowly concentrated around lower values on TREC's 0-3 scale. The right diagram, conversely, exhibits outputs covering a broader range on the target scale, also closely aligned with the diagonal line. Relying solely on ECE values would misleadingly suggest that the first model is better calibrated (ECE: 0.123 < 0.134; lower is better). However, we argue that a model with predictions confined to a limited portion of the desired scale, particularly skewed towards certain values popular in the evaluation set, cannot be considered well-calibrated. CB-ECE offers a solution by re-adjusting the significance of each target scale. In the example shown in Fig.~\ref{figure:reliability-diagram}, CB-ECE more accurately indicates that the second model exhibits better scale calibration (CB-ECE: 1.137 > 1.084; lower is better), which is more in line with our intuitive understanding of model calibration.

\section{Implementation Details} \label{apx:impl}

For all experiments that involve LLM, we employ the LlaMA2-13B-Chat model \citep{touvron_llama_2023}, hosted locally through vLLM\footnote{\url{https://github.com/vllm-project/vllm}} and using quantized weights\footnote{\url{https://huggingface.co/TheBloke/Llama-2-13B-chat-AWQ}}, operated on an A100 GPU. In this study, our neural rankers are based on BERT~\citep{devlin_bert_2019}. We initialize these rankers with weights fine-tuned on MS MARCO, referred to as monoBERT\footnote{\url{https://huggingface.co/veneres/monobert-msmarco}} in Table~\ref{table:main}, or with weights without retrieval-oriented fine-tuning, simply denoted as BERT\footnote{\url{https://huggingface.co/bert-base-uncased}}. For fine-tuning on scale calibration data, we use the AdamW optimizer ~\citep{loshchilov_decoupled_2019} with a learning rate of $3e^{-6}$ and effective batch size of 64 query-document pairs or meta NLEs.\footnote{Note that due to larger model size and more complex input representations, the batch size for fine-tuning neural ranking models is much smaller that training LTR models.} The training is conducted over a maximum of 10 epochs, selecting the best model based on validation set loss. To mitigate the impact of randomness due to the limited dataset size in terms of query numbers, each experiment is run with 5 different random seeds. The metrics reported are averaged across these five runs and the statistical significance is determined using t-tests with Bonferroni correction~\citep{weisstein_bonferroni_nodate} at the 95\% confidence level. 

In relation to the components and hyper-parameters used in Algorithm~\ref{algorithm:aggregation}, following~\citet{quach_conformal_2023}, we employ ROUGE-L~\citep{lin_rouge_2004} as the text similarity function $\mathcal{S}$, set the similarity threshold $\lambda$ to 0.35, the sampling budget $k_l$ (maximum number of responses) to 20 and $k_s$ (maximum number of sentences in the meta NLE) to 30. 
% The repository containing the source code, processed data, and detailed instructions to facilitate the reproduction of the results reported in this study is available at \url{https://github.com/PxYu/LLM-NLE-for-Calibration}.

We use PyGaggle\footnote{\url{https://github.com/castorini/pygaggle}} for acquiring the ranking scores from pre-trained neural ranking models, which are then used for the post-hoc calibration baseline. We use WIG and NQC's implementations in QPP4CS~\citep{meng_query_2023}\footnote{\url{https://github.com/ChuanMeng/QPP4CS}} for the downstream evaluation of QPP.

\section{Case Study} \label{apx:demo}

\begin{table*}[th]
\centering
\begin{tabular}{p{2.5cm}|p{10cm}|p{1cm}}
\toprule
\multicolumn{3}{p{14cm}}{\textbf{Query:} ``Example of what a family advocate does'' \newline \textbf{Document:} ``This person advocates for the best interest of the child, and is tasked with investigating the family situation and advising the court what custody situation would be in the best interests of the child.'' \newline \textbf{Label:} 2 (Highly relevant: The passage has some answer for the query, but the answer may be a bit unclear, or hidden amongst extraneous information).} \\ 
\midrule
\centering \textbf{Approach} & \centering \textbf{Explanation} & \textbf{Score} \\ \hline

\centering Uncalibrated monoBERT & \centering \footnotesize{N/A} & 0.872 \\ \hline
\centering Calibrated monoBERT & \centering \footnotesize{N/A} & 0.397 \\ \hline
\centering Literal Explanation & \footnotesize{here's my answer: relevant the query asks for an example of what a family advocate does, and the document provides an example of a family advocate's role, specifically advocating for the best interests of the child and investigating the family situation to advise the court on the appropriate custody arrangement. this is directly related to the query and the document provides relevant information. the document provides a clear example of what a family advocate does, making it relevant to the query. the document directly addresses the query's question, making it a relevant match. sure, i can help you with that!} & \bf 1.980 \\ \hline
\centering Conditional Explanation & \footnotesize{\textbf{Relevance:} the query ``example of what a family advocate does'' is relevant to the provided document because the document describes the role of a family advocate, which is to advocate for the best interests of the child in family law cases. 
% the document mentions that the advocate investigates the family situation and advises the court on what custody situation would be in the best interests of the child. 
......
the document highlights the key aspects of a family advocate this is an example of what a family advocate might do in practice. this connection between the query and document makes them relevant to each other. 3. focus on the child's best interests: the document emphasizes the importance of prioritizing the child's best interests when making decisions about child custody.}
% sure, i'd be happy to help! 
\newline \newline \footnotesize{\textbf{Non-Relevance:} the query ``example of what a family advocate does'' is not relevant to the document provided because the document does not provide an example of what a family advocate does, but rather describes the responsibilities of a family advocate. the document states that a family advocate advocates for the best interests of the child and investigates the family situation to advise the court on the custody situation that would be in the best interests of the child. this is a description of the role and responsibilities of a family advocate, but it does not provide an example of what a family advocate does in a specific situation.}
% therefore, the query ```example of what a family advocate does'' is not relevant to the document provided. 
& \bf 2.053 \\ \hline
\end{tabular}
\caption{An example showing natural language explanations generated by the LLM lead to better calibrated scores.}
\label{table:case_study}
\end{table*}

We randomly show a pair of query and document from TREC, as well as LLM-generated explanations for them in order to predict calibrated ranking scores. The example is shown in Table~\ref{table:case_study}. The superior effectiveness of our methods in assessing the relevance of the query ``example of what a family advocate does'' to a specific document can be attributed to several aspects. Firstly, the use of NLEs generated by an LLM provides a deeper contextual understanding. For instance, the first approach's explanation elucidates the direct relevance by highlighting how the document exemplifies the role of a family advocate, particularly in advocating for a child's best interests. This detailed breakdown goes beyond mere textual similarity, offering a richer interpretation of the document's relevance to the query. Furthermore, our second approach's dual perspective in explaining both relevance and non-relevance brings a comprehensive understanding. It captures the essence of the document’s role description and contrasts it with the lack of specific examples, thereby offering a deeper analysis of the document-query relationship. By feeding these elaborate explanations into BERT, our methods effectively leverage more informative data, allowing for a more refined and accurate relevance assessment. Consequently, our methods demonstrate a notable improvement in capturing the subtleties and depths of relevance, as evidenced by the scores closer to the label compared to the baseline approaches.

\end{document}